\documentclass[twocolumn,showpacs,amsmath,amssymb,superscriptaddress]{revtex4}
\input{psfig.sty}
\begin{document}
\draft
\title{Anderson localization of polaron states}

\author{Franz X.~Bronold}
\affiliation{Institut f\"ur Theoretische Physik, 
Otto-von-Guericke-Universit\"at Magdeburg,
D-39016 Magdeburg,  Germany} 
\affiliation{Physikalisches Institut, Universit\"a{}t Bayreuth,
D-95440 Bayreuth, Germany}    
\author{Holger Fehske} 
\affiliation{Institut f\"ur Physik, Ernst-Moritz-Arndt Universit\"at
Greifswald, D-17487 Greifswald, Germany}
\affiliation{Physikalisches Institut, Universit\"a{}t Bayreuth, 
D-95440 Bayreuth, Germany}
\date{\today}

\begin{abstract}
Using the vanishing of the typical polaron tunneling rate
as an indicator of the breakdown of itinerancy, we study 
the localization of polaron states in a generic 
model for a disordered polaronic material. 
We find that extremely small disorder causes an Anderson localization
of small polaron states. However, the ratio between the critical disorder 
strength needed to localize all states in the polaron band 
and the renormalized bandwidth is not necessarily smaller than
for a bare electron. 
\end{abstract}
\pacs{71.38.-k, 72.10.Di, 71.35.Aa}
\maketitle
The polaron concept has played a central role in the analysis
of strongly coupled electron-phonon (EP) systems since the pioneering
work of Landau.~\cite{La33} Nevertheless, even fundamental
properties of a polaron are not yet completely understood, especially
with respect to the effect of disorder. Polaron signatures
are currently discussed in compounds such as the colossal magnetoresistance 
manganites~\cite{Mi98}, 
the charge ordered nickelates and bismuthates~\cite{SAL95}, 
and molecular organic semiconductors~\cite{SKB01}
--all {\it complex} materials,
where imperfections potentially play an important role.

The essence of polaron formation is the
crossover from a weakly (phonon) dressed charge carrier to a strongly
mass-enhanced, less mobile polaronic quasiparticle with increasing
EP coupling strength.~\cite{Fi75,dRL83}
This process is usually called {\it self-trapping},
since the distortion that tends to bind the carrier in a deformable lattice
is induced by the carrier itself, i.e., the trapping potential
depends on the state of the carrier.~\cite{Ra82}
Nonetheless, even if the carrier is
``confined'' to a single lattice site
(small polaron), this type of trapping {\it does not imply
localization}. The tunneling between
different lattice sites is still relevant and
a self-trapped carrier resides in an itinerant polaron state.~\cite{dRL83}
On the other hand, it is well known
that for sufficiently strong
randomness the itinerancy of (single electron)
states is destroyed and the states are Anderson localized.~\cite{A58,T70}
Obviously, disorder will 
affect the itinerancy of polaron states. Likewise, it was 
suggested that the ``mobility edge'',
separating itinerant
(extended) and localized states, will be strongly influenced by the
EP interaction~\cite{An72CES83}. At the moment, however, despite
its apparent relevance for various transport processes in
polaronic materials,
little is known rigorously about how
Anderson localization affects polaron states, and vice versa.

As a first step toward  addressing this problem, in Ref.~\cite{BSB01}
a single polaron in a Holstein model with site-diagonal,
binary-alloy-type disorder was studied within the dynamical
mean field approximation (DMFA).~\cite{GKKR96} The DMFA, however, cannot 
(fully) discriminate between itinerant and localized states 
(as shown below), mainly because the randomness is treated at the level of 
the coherent potential approximation.~\cite{EKL74} 

The purpose of this paper is to remedy this shortcoming, and 
to present an advanced theory that is capable of describing both
the Anderson localization and self-trapping phenomena. 
As a generic model we consider the single-particle 
Anderson-Holstein Hamiltonian 
\begin{eqnarray}
{\cal H}&=&\sum_i \epsilon_i n_i 
-J\sum_{\langle ij\rangle} (c_i^\dagger c_j^{} +{\rm H.c.}) 
\nonumber\\
&&-\sqrt{E_p\Omega}\sum_i (b_i + b_i^\dagger) n_i
+ \Omega \sum_i b_i^\dagger b_i,
\label{model}
\end{eqnarray}  
where $J$ is the electron transfer amplitude between neighboring sites, 
$\Omega$ denotes the bare frequency of the optical phonon ($\hbar=1$),
and $E_p$ is the polaron shift.
In Eq.~(\ref{model}), the on-site energies $\{\epsilon_i\}$ are assumed to be
independent, identically distributed random variables 
with a probability density 
$p(\epsilon_i)=(1/\gamma)\theta(\gamma/2-|\epsilon_i|)$. 

Without disorder, the physical properties of 
the Holstein model are governed by three parameter ratios:
the adiabaticity $\alpha=\Omega/J$ [determining whether
the electron is light ($\alpha\ll 1$) or heavy 
($\alpha\gg 1$)~(Ref.~\cite{Ra82})], 
and two dimensionless EP coupling constants  $\lambda=E_p/2J$
and $g^2=E_p/\Omega$.  Extensive numerical studies of the  
Holstein model on finite one-dimensional (1D) and 2D lattices have shown 
that polaron formation sets in if two conditions  
$\lambda\gtrsim 1$ and $g^2 \gtrsim 1$  are 
fulfilled.~\cite{St96,FLW97,BTB99} 
The internal structure of 
the polaron depends on $\lambda$ and $\alpha$. As a result, disorder 
affects polaron states quite differently in the 
adiabatic ($\alpha \ll 1$), nonadiabatic ($\alpha \sim 1$), 
and antiadiabatic ($\alpha \gg 1$) cases. 
Furthermore, in the crossover region, where polaron formation 
sets in, long-range tunneling induced by the EP coupling~\cite{FLW97}  
is expected to decrease the tendency toward  localization. 

To analyze the effects of disorder on polaron formation quantitatively,  
we adopt the recently developed statistical DMFA 
(statDMFA)~(Ref.~\cite{DK97})
to the Anderson-Holstein model. 
This method is an extension of the self-consistent theory of localization 
of bare electrons~\cite{AAT73} to models with interactions, which 
has been successfully applied, e.g., to the disordered 
Hubbard~\cite{DK97} and Anderson lattice~\cite{MD01} 
models. The statDMFA is essentially a 
probabilistic method, based on the self-consistent 
construction of random samples for the (local) physical 
quantities of interest. 

The statDMFA is conveniently constructed for a Bethe lattice 
with finite connectivity $K$.
For later purposes, 
we rescale the transfer amplitude $J\to\tilde{J}/\sqrt{K}$ and
define $\tilde{\lambda}=E_p/2\tilde{J}$ and 
$\tilde{\alpha}=\Omega/\tilde{J}$.
Applying the statDMFA~(Ref.~\cite{DK97}) to  
model~(\ref{model}), the local zero-temperature 
single-electron Green function reads 
${\cal G}_i(z)=[z-\epsilon_i-H_i(z)-{\Sigma}_i(z)]^{-1}$
($z=\omega+i\eta$), with the EP self-energy $\Sigma_i(z)$
and the hybridization function 
\begin{eqnarray}
H_i(z)={{\tilde{J}^2}\over K}\sum_{j=1}^K{1\over{z-\epsilon_j-H_j(z)-\Sigma_j(z)}}.
\label{Hi}
\end{eqnarray}
In the spirit of the self-consistent theory of 
localization,~\cite{AAT73} the statDMFA ignores 
that the functions on the right-hand side 
of Eq.~(\ref{Hi}) should be calculated
for the Bethe lattice with the site $i$ removed. It furthermore 
takes $K$ as the typical number of terms,
although for the initial site there would be $K+1$ terms.
As a consequence, the bare local density of states (LDOS) 
becomes semielliptic (with a half-band-width $\tilde{W}_0/2=2\tilde{J}$), 
irrespective  of the value of $K$, which makes the mean field character 
of the statDMFA evident.~\cite{1D}

The most severe approximation within the statDMFA 
is related to the calculation of $\Sigma_i(z)$. 
Although the connectivity 
of the Bethe lattice is finite in Eq.~(\ref{Hi}),
the self-energy is determined as if the Bethe lattice had  
infinite connectivity.~\cite{remark} 
The self-energy is then
local (as in the DMFA),~\cite{BSB01,S74CdPFF97}
and reads 
$\Sigma_i(z) = E_p 1 \Omega/([F_i^{(1)}(z)]^{-1}- E_p 2\Omega/...)$ 
with $[F_i^{(p)}(z)]^{-1}=z-p\Omega-\epsilon_i-H_{i}^{(p)}(z)$ and 
$H_i^{(p)}(z)=H_i(z-p\Omega)$. Here the energy shift
keeps track of the number of virtual phonons, and the index $p$ is
a positive integer, because of the absence of thermally 
excited phonons. 

A natural measure of the itinerancy of a polaron 
state is the tunneling rate from a given site
defined as the imaginary part of the hybridization
function $\Gamma_i(\omega)=(\pi \tilde{J}^2/K)\sum_{j=1}^K N_j(\omega)$, 
where $N_j(\omega)=-(1/\pi){\rm Im} {\cal G}_j(\omega)$ is the LDOS. 
A vanishing tunneling rate $\Gamma_i(\omega)$ implies a localized state 
at energy~$\omega$. 

Due to the randomness in the on-site energies,
the tunneling rate $\Gamma_i(\omega)$ is a random variable,
and the question of whether it vanishes or not depends on 
the probability density exhibiting, quite generally, characteristic 
differences  
for itinerant and localized states.~\cite{A58,AAT73} 
For an itinerant 
state, the probability density is narrow and (approximately) symmetric. 
Thus, the most probable value and the arithmetic mean 
(average), ${\Gamma}^{\rm av}(\omega)=(1/N)\sum_i^{N}\Gamma_i(\omega)$, 
$N$ being the total number of lattice sites, are almost identical and
${\Gamma}^{\rm av}(\omega)$ will be a good estimate for the tunneling rate. 
On the other hand, for a localized state, the probability density is 
very wide with an extremely long tail toward large values,
implying that the most probable value 
is much smaller than the average.
In that case, the geometric mean (typical value) 
\begin{equation}
\Gamma^{\rm typ}(\omega)=\exp 
\left[\frac{1}{N}\sum_i^{N}\log\Gamma_i(\omega)\right] 
\end{equation}
seems to be a more appropriate estimate of the 
tunneling rate. Accordingly, a vanishing {\it typical} 
(rather than average) tunneling rate  
should be taken as a localization criterion.~\cite{AAT73}   

The typical tunneling rate is most directly 
calculated from a recursion scheme for the hybridization
function 
\begin{eqnarray}
H_{i}^{(p)}={\tilde{J}^2\over K}\sum_{l=1}^K
   \frac{1}{\displaystyle [F_l^{(p)}]^{-1}-
   \frac{E_p 1 \Omega}{\displaystyle [F_l^{(p+1)}]^{-1}-
   \frac{E_p 2 \Omega}{...}}}.
\label{RS}
\end{eqnarray}
Demanding that $H_i^{(p)}(z)$ are independent, 
identically distributed random variables, Eq.~(\ref{RS})
can be used to construct a random sample $\{H_i^{(p)}(z)\}$
($i=1,...,N$ and $p=0,1,...,M$) for the
hybridization function, starting from an initial random
configuration which is successively updated with a 
sampling technique similar to the one described in 
Ref.~\cite{AAT73}. More precisely, we are constructing a self-consistent
random sample because the probability density for 
$H_i^{(p)}(z)$ is assumed to be the same on both sides
of Eq. (\ref{RS}).  
Clearly, the larger $N$ the better the statistics. 
A correct description of the polaron effect 
requires the cutoff $M$ for the continued fraction to be much larger 
than the average number of phonons bound in
the polaron. Specifically, for the results presented in
this paper $N=50\,000$ and $M=35$. 

It is instructive to consider Eq.~(\ref{RS}) for $K=\infty$. 
Applying the law of large numbers shows that  
$H_i^{(p)}(z)$ can be replaced by its average 
${H}^{{\rm av}(p)}(z)$. Then, $(1/K)\sum_l^K(...)$ denotes 
the (sample) average solely over $\epsilon_l$, and Eq.~(\ref{RS})
reduces to the DMFA equation for ${H}^{{\rm av}(p)}(z)$. 
(The scaling of $J$ makes this point particularly clear.)
Obviously, the tunneling rate is then $\pi\tilde{J}^2{N}^{\rm av}(z)$, 
i.e., it is proportional to the average LDOS,
which is always finite, irrespective of
whether the state at energy $\omega$ is 
itinerant or localized.~\cite{A58,T70} Thus, 
for $K=\infty$, i.e., in the DMFA, itinerant and localized states cannot 
be distinguished.
\begin{figure}[t]
\psfig{figure=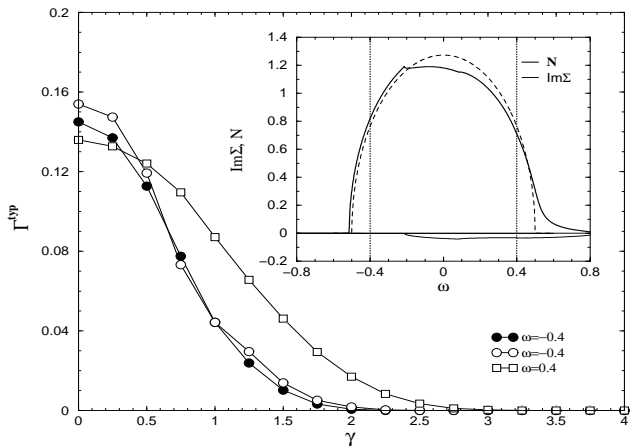,height=6.0cm,width=0.95\linewidth,angle=-90}
\caption[fig1]
{Typical tunneling rates $\Gamma^{\rm typ}(\omega;\gamma)$ in the
(non-to-antiadiabatic) weak EP coupling regime
($\tilde{W}_0=1.0$, $\tilde{\alpha}=1.2$,
and $\tilde{\lambda}=0.067$). Filled circles denote data
for $\tilde{\lambda}=0$. The inset shows the LDOS
$N(\omega)$ and the
imaginary part of the EP self-energy
${\rm Im}\Sigma(\omega)$ for $\gamma=0$; dashed
and (vertical) dotted lines indicate, respectively, the density of
states for $\tilde{\lambda}=0$ and the
energies $\omega$ for which $\Gamma^{\rm typ}(\omega;\gamma)$ is
plotted.}
\label{fig1}
\end{figure}
\begin{figure}[h]
\psfig{figure=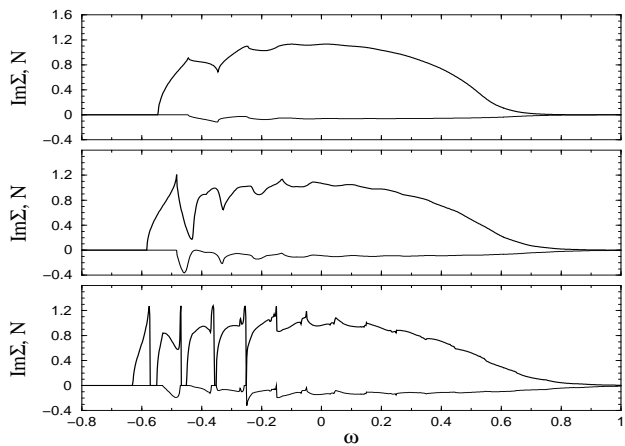,height=6.0cm,width=0.95\linewidth,angle=-90}
\caption[fig2]
{Evolution of the LDOS
$N(\omega)$ (thick line) and the imaginary part of
the EP self-energy ${\rm Im}\Sigma(\omega)$ (thin
line) for the ordered Holstein model ($\gamma=0$) with
increasing EP coupling $\tilde{\lambda}=0.4, 0.7,$ and $1.0$
(from top to bottom). Results are given for $\tilde{\alpha}=0.4$
and $\tilde{W}_0=1.0$.}
\label{fig2}
\end{figure}  

In the following we present representative numerical results for $K=2$ and
$\eta=10^{-8}$. We start our discussion with the 
(nonadiabatic-to-antiadiabatic) 
weak EP coupling regime. 
Figure \ref{fig1} displays typical tunneling rates  
$\Gamma^{\rm typ}(\omega;\gamma)$ as a function of disorder strength $\gamma$
for two energies $\omega$, one below and one above 
the optical-phonon emission threshold, which, due to the large value 
of $\tilde{\alpha}$, lies far inside the band (see the inset). Below the threshold, 
${\rm Im}\Sigma(\omega)=0$ and the states are coherent. 
Above the threshold, however, inelastic polaron-phonon scattering yields
${\rm Im}\Sigma(\omega)\neq 0$, i.e., the states are incoherent, implying 
that the quantum interference needed for localization~\cite{LR85} is
significantly suppressed. We therefore expect
the  critical disorder strength $\gamma_c$, for which  
$\Gamma^{\rm typ}(\omega;\gamma)$ vanishes, to be larger above
the phonon emission threshold than below it. Indeed, we find 
$(\gamma_c/\tilde{W}_0)_{\omega=-0.4}\approx 2.25$ and 
$(\gamma_c/\tilde{W}_0)_{\omega=0.4}\approx 3.0$. 
It is rather encouraging that the statDMFA recovers this basic 
feature of the physics of localization.  
Since the EP coupling is weak the states below 
the threshold are basically bare electron states and   
$(\gamma_c/\tilde{W}_0)_{\omega=-0.4}\approx 
(\gamma_c/\tilde{W}_0)_{\omega=-0.4}^{\tilde{\lambda}=0}\approx 2.0$, 
the corresponding ratio for a bare electron.~\cite{gcrit} 

Increasing the EP coupling, polaron formation starts and the
LDOS fragments into an increasing number of (more or less) coherent
subbands.~\cite{S74CdPFF97} Figure  \ref{fig2} illustrates, for
the ordered Holstein model ($\gamma=0$), the evolution of the LDOS with 
increasing EP coupling $\tilde{\lambda}$ for fixed $\tilde{\alpha}$ and $\tilde{W}_0$. 
As far as localization is concerned, the lowest subband is of 
particular interest, because it is completely coherent and 
the quantum interference needed for localization is not 
masked by inelastic polaron-phonon scattering.
\begin{figure}[t]
\psfig{figure=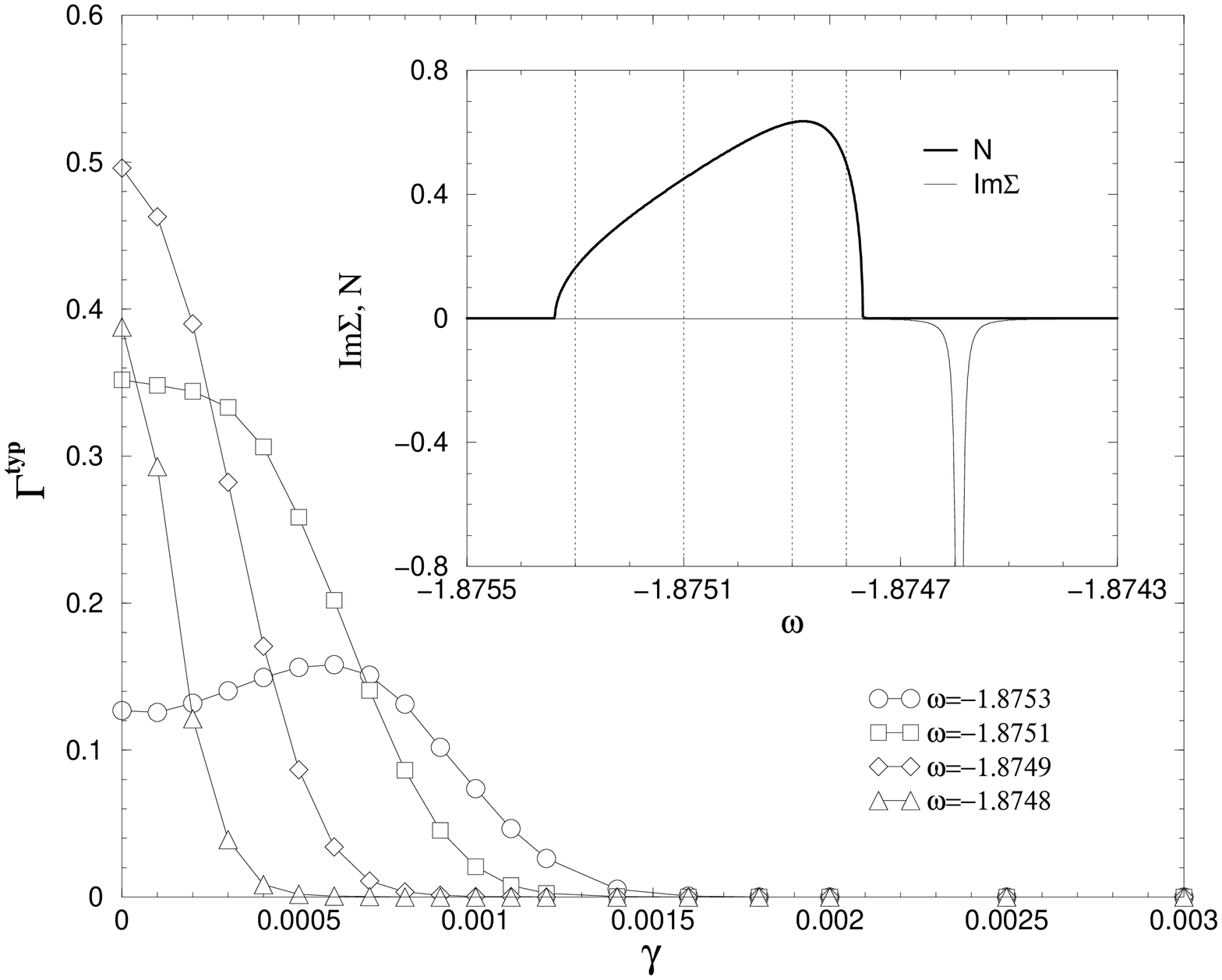,height=6.0cm,width=0.95\linewidth,angle=-90}
\psfig{figure=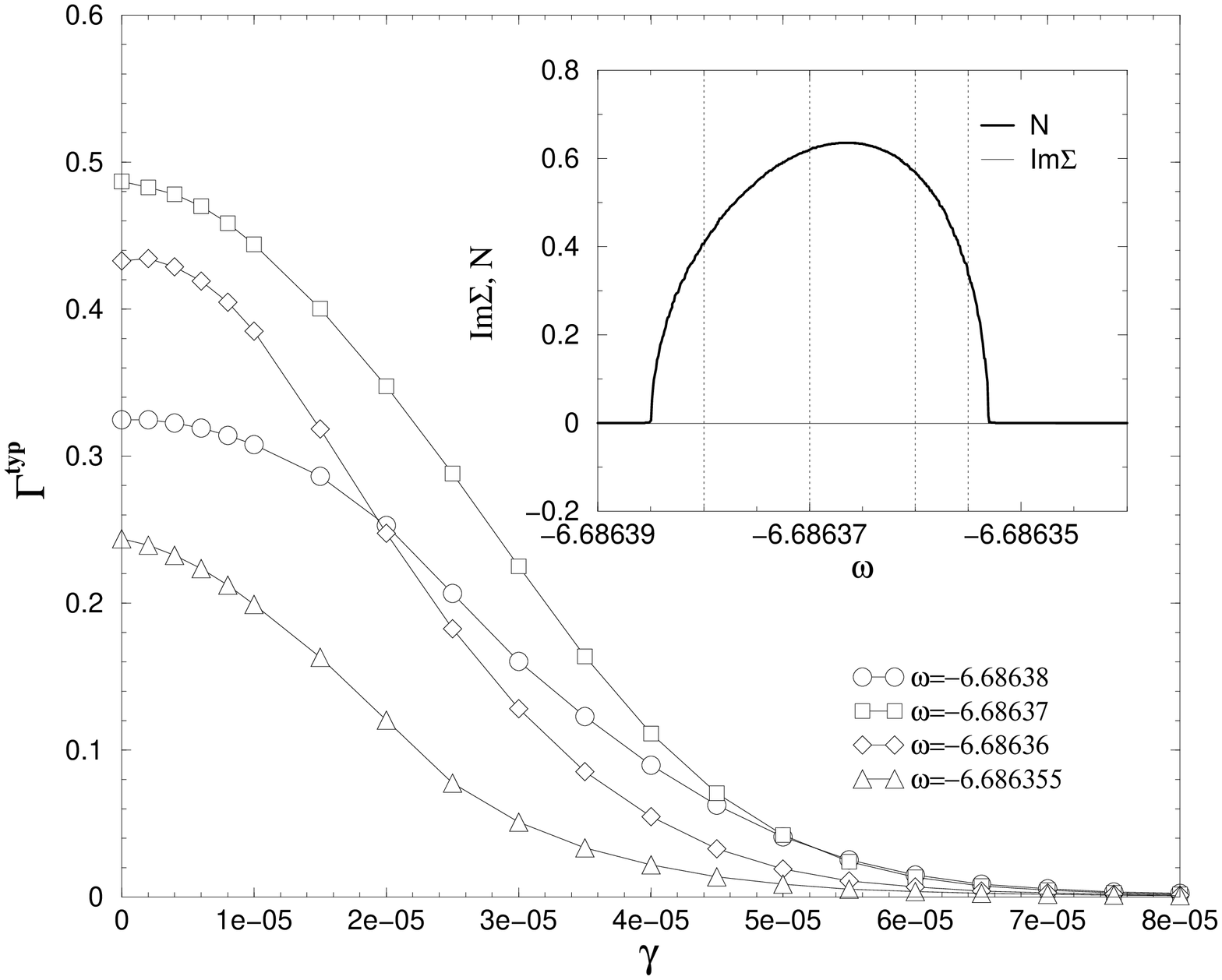,height=6.0cm,width=0.95\linewidth,angle=-90}
\caption[fig6]
{The upper and lower panels show, respectively, typical tunneling rates
$\Gamma^{\rm typ}(\omega;\gamma)$ in the (adiabatic) intermediate
($\tilde{\alpha}=0.4$, $\tilde{\lambda}=1.8$) and (non-to-antiadiabatic)
strong EP coupling regime
($\tilde{\alpha}=1.2$ and $\tilde{\lambda}=6.6\bar{6}$)
using $\tilde{W}_0=2.0$.
The insets display the subband LDOS $N(\omega)$ and the
imaginary part of the EP self-energy ${\rm Im}\Sigma(\omega)$
for $\gamma=0$ (vertical dotted lines indicate
the energies $\omega$ for which $\Gamma^{\rm typ}(\omega;\gamma)$ is
plotted). Note the small spectral weight of the
lowest polaron subband.}
\label{fig6}
\end{figure}      
Numerical studies of the ordered Holstein model demonstrated that 
the band dispersion of 
the lowest subband depends on both $\tilde{\lambda}$ and
$\tilde{\alpha}$, and may differ 
significantly from the rescaled bare band.~\cite{St96,FLW97}
Specifically, the states at the bottom of the subband 
are mostly electronic and rather mobile due to long-range tunneling 
induced by EP coupling, whereas the states at the top 
of the subband are rather phononic and immobile.~\cite{FLW97} 
Therefore, we expect that 
the critical disorder strength 
needed to localize states at the top of the subband is 
much smaller than for states at the bottom. This behavior should
be most pronounced in the (adiabatic) intermediate EP coupling regime. 

The upper panel of 
Fig.~\ref{fig6}, showing typical tunneling rates 
$\Gamma^{\rm typ}(\omega;\gamma)$ 
as a function of 
disorder strength $\gamma$ for the lowest polaron subband, 
impressively confirms this striking behavior.
The strong asymmetry of the LDOS (see the inset) is a 
direct consequence of the ``hybridization'' with the (optical)
phonon branch, leading to a band flattening at the 
Brillouin-zone boundary. 
As can be seen, states at the zone boundary are 
very susceptible to disorder and
the critical disorder strength $\gamma_c$ for which 
$\Gamma^{\rm typ}(\omega;\gamma)$
vanishes is substantially smaller than  
at the bottom of the subband. Clearly, the absolute scale of 
disorder affecting the subband  is on the order
of the subband width $\tilde{W}_1$, which is strongly renormalized in comparison
with the bare band width $\tilde{W}_0$. 
Different from the pure Anderson model,
the critical disorder strength needed to localize {\it all} states of the 
subband is determined by the states at the  bottom (not at the center). 
We find $(\gamma_c/\tilde{W}_1)_{\rm complete}\approx 2.8$, which 
is larger than the corresponding ratio for a bare electron 
[$(\gamma_c/\tilde{W}_0)_{\rm complete}\approx 2.25$] because of the long-range 
tunneling induced by the EP coupling. Thus,   
contrary to naive expectations, in the intermediate-coupling 
(crossover) regime,
a polaron is more difficult 
to localize than a bare electron.

Finally, in the lower panel of Fig.~\ref{fig6}, we present typical tunneling 
rates $\Gamma^{\rm typ}(\omega;\gamma)$ in the 
(nonadiabatic-to-antiadiabatic) strong EP
coupling regime. The phonon admixture of the polaron states within 
the lowest subband is now nearly energy independent.
Concomitantly, long-range tunneling as well as 
band flattening are negligible and the LDOS is rather 
symmetric (see the inset).~\cite{FLW97}  
Accordingly, disorder should 
affect all the states within the subband more or less
equally. Indeed, the critical disorder strength 
$\gamma_c$ for which $\Gamma^{\rm typ}(\omega;\gamma)$ 
vanishes is now almost energy independent.  
We find $(\gamma_c/\tilde{W}_1)_{\rm complete} \approx 2.5$,
which is smaller than the ratio for intermediate
EP coupling, indicating that in the extreme strong EP coupling 
adiabatic ($\tilde{\lambda}\gg 1$) and 
nonadiabatic-to-antiadiabatic
($g^2\gg 1$) regimes
$(\gamma_c/\tilde{W}_1)_{\rm complete}$ approaches the corresponding
ratio for a bare electron.
Here, the band collapse 
does only change the overall energy scales.

To summarize, we presented, as a first step toward a
transport theory for disordered polaronic materials,
a theoretical approach capable
of describing polaron formation 
and Anderson localization. 
We showed that the typical tunneling
rate from a given site is an appropriate quantity in order
to distinguish itinerant and localized polaron states.
Focusing on the  disorder-induced localization of a single polaron
in the intermediate-to-strong EP coupling regime, we found
that extremely small material imperfections turn
itinerant polaron states into localized polaron
states, suggesting that in real materials (small)
polarons are presumably always localized. In the crossover regime,
the critical disorder strength, in units of the subband width,
needed to localize all states of the lowest polaron subband
can be larger than that for the pure Anderson model. From this
perspective, disorder-induced localization of a polaron
is not necessarily easier than localization of a bare electron.



\begin{references}
\bibitem{1D}
The statDMFA should be considered
as the high-dimensional limit of a localization theory for
interacting electrons and obviously gives erroneous results
for one dimension.
\bibitem{remark}
The statDMFA therefore does not treat disorder 
and EP interaction on an equal footing. To go beyond
it would, however, render the calculations intractable.
\bibitem{gcrit}
That is, 
$(\gamma_c/J)_{\omega=-0.4}^{\tilde{\lambda}=0}\approx 11$,
consistent with the results given in Ref. \cite{AAT73}.
\end{references}
\end{document}